\documentclass[a4paper,11pt]{article}
\usepackage{pos}

\title{Time, quantum entanglement, and particle decay}

\author*[a]{J. A. Aguilar-Saavedra}

\affiliation[a]{Instituto de F\'isica Te\'orica IFT-UAM/CSIC \\ c/Nicol\'as Cabrera 13--15, 28049 Madrid, Spain}

\emailAdd{ja.a.s@csic.es}

\abstract{We investigate the role of time ordering in entanglement experiments involving unstable particles, focusing on $\mu^+ \mu^-$ pairs produced in a maximally-entangled spin state. We analyse the correlations between measurements performed by two experimenters, Alice (who measures $\mu^-$ spin) and Bob (who measures $\mu^+$ decay products). Remarkably, the observed correlations persist irrespective of whether Bob's muon decays before or after Alice's spin measurement. We further discuss different interpretations of the same empirical results depending on the observer's reference frame. The findings reinforce the viewpoint that the Copenhagen interpretation of measurement is a mathematical tool rather than a literal account of physical reality.}

\FullConference{
Proceedings of the Corfu Summer Institute 2025 "School and Workshops on Elementary Particle Physics and Gravity" (CORFU2025) 27 Apr - 04 May, and 24 August - 28 September, 2025, Corfu, Greece
}

\begin{document}
\maketitle

\section{Introduction}

Entanglement experiments at colliders introduce a key feature absent from the traditional setups with electrons or photons: particle decay. As is well known, the decay makes it impossible to directly measure the particle spin; instead, only expectation values of spin operators can be inferred from the angular distributions of the decay products. This limitation effectively precludes Bell-type tests of quantum mechanics (QM) against hidden-variable theories. But the very presence of a decay also highlights the role of time in entanglement experiments. This perspective has been explored for kaon pairs, focusing on flavour entanglement~\cite{Bernabeu:2019gjs}, and for spin entanglement of muon pairs~\cite{Aguilar-Saavedra:2023lwb}. In what follows, we build upon the latter case and its interpretation. In particular, we highlight the fact that different observers may arrive at quite different interpretations of the same experimental results.

\section{Experiment setup}

Following Ref.~\cite{Aguilar-Saavedra:2023lwb}, we consider a $\mu^+ \mu^-$ pair in a maximally-entangled Bell state. At particle colliders, several mechanisms can produce muon pairs with entangled spins. The decay of a pseudo-scalar particle, such as the $\eta$ meson, produces a $\mu^+ \mu^-$ pair in a spin singlet $[ |+-\rangle - |-+\rangle] /\sqrt{2}$. Also, a spin-triplet state $[ |+-\rangle + |-+\rangle] /\sqrt{2}$  arises in the low-energy Drell-Yan process $e^+ e^- \to \mu^+ \mu^-$, at production angle $\theta = \pi/2$.

\begin{figure}[htb]
\begin{center}
\includegraphics[height=6cm,clip=]{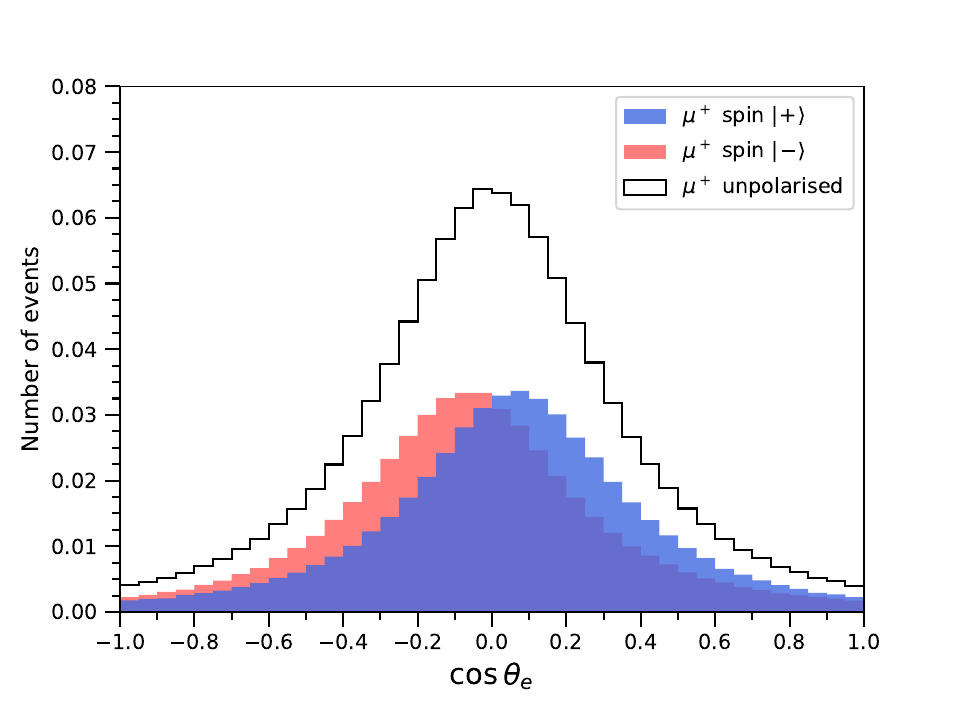} 
\caption{Polar angle distribution in laboratory frame for the $e^+$ resulting from $\mu^+$ decay in spin-up (blue) and spin-down (red) eigenstates. The sum of both corresponds to unpolarised $\mu^+$ decays, and is displayed in white.}
\label{fig:costh}
\end{center}
\end{figure}

Two experimenters, Alice and Bob, perform measurements on $\mu^-$ and $\mu^+$, respectively. Alice measures the $\mu^-$ spin,\footnote{Stern-Gerlach experiments are usually performed with neutral particles, since charged ones are deflected by magnetic fields. It is conceivable that a combination of electric and magnetic fields could allow to measure the spin of charged particles. In any case, this experimental complication is out of the scope of the present discussion.} and Bob lets his muons decay $\mu^+ \to e^+ \nu \nu$, measuring the direction of the outcoming $e^+$. For muon decays, the normalised decay rate in their rest frame is, after integrating over the momenta of the two neutrinos and the $e^\pm$ energy ~\cite{Bouchiat:1957zz}
\begin{equation}
\frac{1}{\Gamma}\frac{d\Gamma}{d\!\cos \theta_e^*} = \frac{1}{2} \left[ 1 + \frac{1}{3} \kappa P_{\hat n} \cos \theta_e^* \right] \,.
\label{ec:mudec1D}
\end{equation}
where $\theta_e^* = \angle(\vec p_{e}^*,\hat n)$ is the angle between the $e^\pm$ three-momentum (we denote quantities in the muon rest frame with an asterisk) and an arbitrary direction $\hat n$; $P_{\hat n}$ is the muon polarisation in that axis, $P_{\hat n}= 2 \langle \vec S \cdot \hat n \rangle$; $\kappa = 1$ for $\mu^+$ decays and $\kappa = -1$ for $\mu^-$ decays. This distribution could in principle be used to determine expected values of spin operators; however, the muon rest frame cannot be experimentally reconstructed because of the two missing neutrinos. Still, the polarisations in directions orthogonal to the momentum can be measured. Let us choose the $\hat x$ axis along the $\mu^+$ momentum direction in laboratory frame. Then, for example, the up-down asymmetry
\begin{equation}
\int_0^1 \frac{1}{\Gamma}\frac{d\Gamma}{d\!\cos \theta_{e}^*}
- \int_{-1}^0 \frac{1}{\Gamma}\frac{d\Gamma}{d\!\cos \theta_{e}^*} = \frac{P_3}{6} 
\label{ec:AP}
\end{equation}
provides the polarisation in the $\hat z$ axis, $P_3$. The asymmetry is the same when evaluated in the laboratory frame in terms of the angle $\theta_{e}$, because the $z$ component of the $e^+$ momentum is not changed by the boost. Note, however, that the $\cos \theta_{e}^*$ and $\cos \theta_{e}$ distributions are different, the latter depending on the boost. For a $\mu^+ \mu^-$ pair resulting from $\eta$ decay, the distributions are presented in Fig.~\ref{fig:costh}.
Likewise, the polarisation in any axis $\hat n$ within the $yz$ plane can be measured. Taking $\theta_{e}$ as the angle $\angle(\vec p_{e},\hat n)$, the distributions for the spin-up and spin-down states $|\!\uparrow\rangle_{\hat n}$, $|\!\downarrow\rangle_{\hat n}$ are the same as the blue and red ones in Fig.~\ref{fig:costh}, respectively, and the asymmetry around $\cos \theta_e = 0$ equals $P_{\hat n}/6$.

The experiment then proceeds as follows. A muon pair is produced from a source at $t=0$. Alice measures the $\mu^-$ spin in her direction of choice (contained in the $yz$ plane) at time $t_1$. Bob records the direction of $e^+$ from $\mu^+$ that decays at time $t_2$, not communicating with Alice.\footnote{In a practical implementation, the time ordering $t_1 < t_2$ or $t_1 > t_2$ can be controlled by the distance of Alice and Bob to the source. Also, if the $\mu^+$ does not decay in Bob's detector, the muon pair is discarded from the analysis.} This is repeated for several muon pairs and, after that, Alice and Bob compare their results. Notice that Bob does not have any measurement choice as Alice does; he records $\mu^+$ decays and is able to later determine $P_{\hat n}= 2 \langle \vec S \cdot \hat n \rangle$ for any $\hat n$ in the $yz$ plane, using the corresponding $\cos \theta_{e}$ distribution measured from a set of events.

\section{Correlations for $t_1 < t_2$}
\label{sec:3}

We first address the simplest situation: Alice measures the spin of $\mu^-$ and Bob records $\mu^+$ decays at a later time $t_2 > t_1$. For simplicity, we work with $\mu^+ \mu^-$ pairs in spin-singlet state, as resulting from $\eta$ decay. 

Alice decides to measure $\mu^-$ spins in the $\hat z$ direction, for example.  Her measurement projects the spin state of the muon pair into $|+-\rangle$ or $|-+\rangle$, depending on the outcome. Therefore, Bob's muons are in a $S_3$ spin eigenstate when they decay. After Alice and Bob perform their measurements for a number $N$ of muon pairs, they can communicate their results. If Bob selects from his sample the $\mu^+$ for which Alice measured a spin-up state $|+\rangle$, he finds that the $\cos \theta_e$ distribution in this sub-sample corresponds to a pure spin-down state, $P_3 = -1$. Conversely, if Bob selects the $\mu^+$ for which Alice measured a spin-down state $|-\rangle$, he finds a $\cos \theta_e$ distribution that agrees with a pure spin-up state, $P_3 = 1$.

Analogous results are found if Alice measures spin in an arbitrary direction $\hat n$ in the $yz$ plane, due to the spherical symmetry of the spin-singlet state. When Bob selects from his sample the events for which Alice measured a spin-up state
$|\!\uparrow\rangle_{\hat n}$, the $\cos \theta_e$ distribution agrees with a  spin-down state, $P_{\hat n} = -1$. When he selects the events in which Alice measures $|\!\downarrow\rangle_{\hat n}$, his $\cos \theta_e$ distribution is the one for $P_{\hat n} = 1$.

In addition, the quantumness of the correlation between Alice and Bob's measurements can be verified using the Clause-Horne-Shimony-Holt (CHSH) inequalities~\cite{Clauser:1969ny}
\begin{equation}
\left| \langle AB \rangle - \langle AB' \rangle + \langle A'B \rangle + \langle A'B' \rangle \right| \leq 2 \,,
\label{ec:CHSH}
\end{equation}
where $A$, $A'$ are spin operators for $\mu^-$ (Alice) and $B$, $B'$ spin operators for $\mu^+$ (Bob), normalised to have eigenvalues $\pm 1$. We can choose
\begin{align}
& A = \sigma_3 \,,\quad A' = \sigma_2 \,,\quad B = \frac{1}{\sqrt 2}(\sigma_2 + \sigma_3) \,,\quad
B' = \frac{1}{\sqrt 2}(\sigma_2 - \sigma_3) \,,
\end{align}
with $\sigma_i = 2 S_i$ the Pauli matrices. The expected values are computed by weighing 
$\langle B \rangle$ and $\langle B' \rangle$ with the eigenvalue of $A$ or $A'$.\footnote{In the $yz$ plane, $\langle B \rangle$ is measured from the asymmetry along the bisector of the second and fourth quadrants, with the positive direction in the first quadrant. Likewise, $\langle B' \rangle$ is measured from the asymmetry along the bisector of the first and third quadrants, with the positive direction in the fourth quadrant.} For example,
\begin{equation}
\langle AB \rangle = \frac{1}{2} \left[ \langle B \rangle_\uparrow - \langle B \rangle_\downarrow \right] \,,
\end{equation}
where the up (down) arrows indicate that the averages are restricted to events with positive (negative) eigenvalue obtained in the measurement of $S_3$ for the negative muon,
\begin{equation}
\langle B \rangle_\uparrow = - \langle B \rangle_\downarrow = - \frac{1}{\sqrt{2}} \,.
\end{equation}
The other expected values are calculated similarly, and one has
\begin{equation}
\langle AB \rangle =  - \langle AB' \rangle = \langle A'B \rangle = \langle A'B' \rangle = -\frac{1}{\sqrt 2} \,,
\end{equation}
so the l.h.s. of (\ref{ec:CHSH}) equals $2\sqrt 2$, maximally violating the inequality. 

So far, the correlations between the measurements found for $t_1 < t_2$ are just as expected: we start from a maximally-entangled Bell state, which collapses into a spin eigenstate when Alice performs her measurement. What is more striking, however, is that the same correlations between Alice and Bob's measurements persist for $t_1 > t_2$, i.e. when Bob's muon decays before Alice performs her measurement. Intriguingly, in this case Alice can still choose her spin measurement direction after the $\mu^+$ has decayed! The origin of the correlations when $t_1 > t_2$ will be clarified in the next section.

\section{Correlations for $t_1 > t_2$}
\label{sec:4}

\subsection{Canonical perspective}
\label{sec:4.1}

To analyse this case we use the formalism of post-decay density operators in Ref.~\cite{Aguilar-Saavedra:2024fig}. The initial state of the $\mu^+ \mu^-$ pair can be described by an operator with a general form
\begin{equation}
\rho = \sum_{ijkl} \rho_{ij}^{kl} |\phi_i \chi_k \rangle \langle \phi_j \chi_l | \,,  
\end{equation}
where $|\phi_{\pm 1} \rangle$ and $|\chi_{\pm 1} \rangle$ are the $S_3$ spin eigenstates of $\mu^+$ and $\mu^-$, respectively. For the spin-singlet state, the only non-zero elements are
\begin{equation}
\rho_{11}^{-1-1} = \rho_{-1-1}^{11} = - \rho_{1-1}^{-11} = - \rho_{-11}^{1-1} = 1/2 \,.
\label{ec:rhosinglet}
\end{equation}
After the decay of $\mu^+$, the spin state of its decay products and the surviving $\mu^-$ is described by an (unnormalised) density operator
\begin{equation}
\hat \rho^\prime = \sum_{\alpha \beta k l} (M \rho^{kl} M^\dagger)_{\alpha \beta} |\xi_\alpha \chi_k \rangle \langle \xi_\beta \chi_l | \,,
\end{equation}
where $|\xi_\alpha \rangle$ are the spin states of the $\mu^+$ decay products and
\begin{equation}
M_{\alpha j} = \langle P \xi_\alpha | T | \phi_j \rangle
\end{equation}
the decay amplitudes for fixed momenta of the $\mu^+$ decay products, which we generically denote as $|P\rangle$. For the weak interaction involved in the decay $\mu^+ \to e^+ \bar \nu_\mu \nu_e$ there is only one combination $|\xi \rangle$ of helicities giving non-zero amplitures, namely helicities $\lambda = 1/2$ for $e^+$ and $\bar \nu_\mu$ and $\lambda = -1/2$ for $\nu_e$. We therefore drop the subindex $\alpha$ for those states. 
By explicit calculation one obtains
\begin{align}
M_{1} & = 16 G_F \left[ m_\mu E_2^* E_3^* E_4^* \right]^{\frac{1}{2}} e^{i \frac{\varphi_3^*}{2}} \sin \frac{\theta_3^*}{2} \left[
e^{i \frac{\varphi_4^*-\varphi_2^*}{2}} \cos \frac{\theta_4^*}{2} \sin \frac{\theta_2^*}{2} - e^{-i \frac{\varphi_4^*-\varphi_2^*}{2}} \cos \frac{\theta_2^*}{2} \sin \frac{\theta_4^*}{2} \right] \,, \notag \\
M_{-1} & = 16 G_F \left[ m_\mu E_2^* E_3^* E_4^* \right]^{\frac{1}{2}} e^{-i \frac{\varphi_3^*}{2}} \sin \frac{\theta_3^*}{2} \left[
e^{i \frac{\varphi_4^*-\varphi_2^*}{2}} \cos \frac{\theta_4^*}{2} \sin \frac{\theta_2^*}{2} - e^{-i \frac{\varphi_4^*-\varphi_2^*}{2}} \cos \frac{\theta_2^*}{2} \sin \frac{\theta_4^*}{2} \right] \,,
\end{align}
with $G_F$ the Fermi constant. The momenta of $e^+$, $\nu_e$ and $\bar \nu_\mu$ in the $\mu^+$ rest frame are labelled as $p_2^*$, $p_3^*$ and $p_4^*$, respectively, with $p_i^* = (E_i^*, \vec p_i^*)$, and $\vec p_i^*$ having polar and azimuthal angles $(\theta_i^*,\varphi_i^*)$. Therefore, for the initial spin-singlet state with non-zero elements (\ref{ec:rhosinglet}) we have
\begin{equation}
\hat \rho^\prime = \frac{1}{2} \left[
| M_{-1} |^2 | \xi \chi_1 \rangle \langle \xi \chi_1 |
+ | M_{1} |^2 | \xi \chi_{-1} \rangle \langle \xi \chi_{-1} | - M_{-1} M_{1}^* | \xi \chi_1 \rangle \langle \xi \chi_{-1} |
- M_{1} M_{-1}^* | \xi \chi_{-1} \rangle \langle \xi \chi_{1} | \right] \,.
\end{equation}
The trace over spin degrees of freedom of the $\mu^+$ decay products is trivial and yields the spin density operator for $\mu^-$
\begin{equation}
\hat \rho_A^\prime = \frac{1}{2} \left[
| M_{-1} |^2 | \chi_1 \rangle \langle \chi_1 |
+ | M_{1} |^2 | \chi_{-1} \rangle \langle \chi_{-1} | 
- M_{-1} M_{1}^* | \chi_1 \rangle \langle \chi_{-1} | - M_{1} M_{-1}^* | \chi_{-1} \rangle \langle \chi_{1} | \right] \,.
\label{ec:rhoA}
\end{equation}
For fixed momenta $|P\rangle$, this operator corresponds to a pure state $1/\sqrt 2 [ M_{-1} |\chi_1> - M_{1} |\chi_{-1}\rangle ]$. The relative probabilities that Alice obtains spin-up and spin-down when she measures $S_3$ are $| M_{-1} |^2$ and $| M_{1} |^2$, respectively, and they do depend on the $\mu^+$ decay kinematics.

\begin{figure}[htb]
\begin{center}
\includegraphics[height=6cm,clip=]{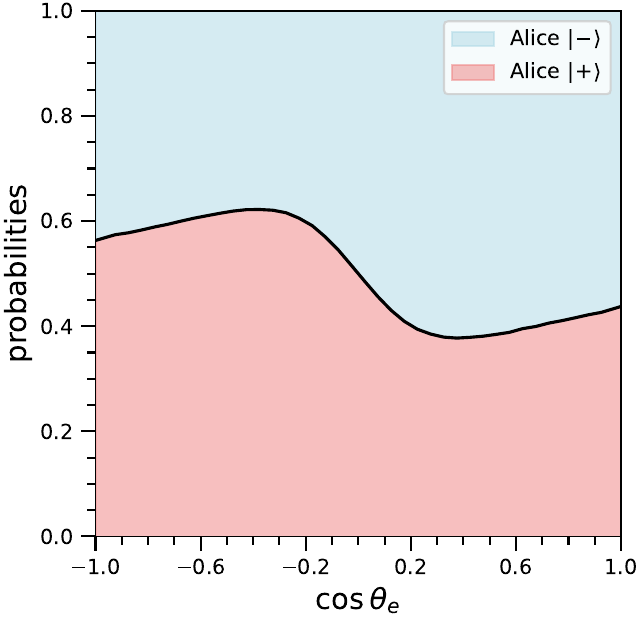} 
\caption{Probabilities of Alice obtaining spin-up and spin-down in her $S_3$ measurement, depending on the laboratory-frame angle $\theta_e$ of the $e^+$ resulting from $\mu^+$ decay.}
\label{fig:Aliceprob}
\end{center}
\end{figure}

When the momenta of $\mu^+$ decay products are not considered fixed, the unnormalised spin density operator for $\mu^-$ is obtained from integration of (\ref{ec:rhoA}). After integrating over the neutrino momenta and the $e^+$ energy in $\mu^+$ rest frame,
one obtains from the coefficients of $| \chi_1 \rangle \langle \chi_1 |$ and $ | \chi_{-1} \rangle \langle \chi_{-1} |$ the relative probabilities for Alice measuring spin-up or spin-down, respectively,
\begin{equation}
P(+) : P(-) = \left( 1 - \frac{1}{3} \cos \theta_2^* \right)  :  \left( 1 + \frac{1}{3} \cos \theta_2^* \right)  \,.
\label{ec:aliceprob}
\end{equation}
One easily recognises here the decay angular distributions for $\mu^+$ in spin eigenstates $S_3 = \mp 1/2$, c.f. (\ref{ec:mudec1D}). As we see, the decay of $\mu^+$ is influencing Alice's measurement in a subtle way:
the more consistent the $\mu^+$ decay is with $S_3 = -1/2$, the more probable is that Alice will get spin-up in her measurement, and the opposite. In terms of the laboratory-frame angle $\theta_e$, the probabilities are given in Fig.~\ref{fig:Aliceprob}. 

Then, when Alice and Bob select the sub-sample for which Alice measures spin-up, the $\cos \theta_e$ distribution for Bob's decay is precisely the red one in Fig.~\ref{fig:costh}, corresponding to $\mu^+$ with spin $|-\rangle$. Conversely, selecting 
the pairs in which Alice measures spin-down, the resulting $\cos \theta_e$ distribution is the blue one, corresponding to spin $|+\rangle$. This end can be numerically verified by performing pseudo-experiments in which we:
\begin{enumerate}
\item consider a random orientation for $e^+$ in the $\mu^+$ rest frame;
\item assign probabilities for Alice's measurement of $S_3$ according to (\ref{ec:aliceprob}) and choose a random outcome;
\item compute $\theta_e$ in laboratory frame.
\end{enumerate}
The $\cos \theta_e$ distributions for the two sub-samples, those in which Alice measures spin-up and spin-down, agree with Bob's muon in spin eigenstates $|-\rangle$ and $|+\rangle$, respectively. Clearly, similar conclusions can be obtained when Alice measures spin in a direction $\hat n$ other than the $\hat z$ axis.

\subsection{Unorthodox view}

Alternatively, one may consider that Alice's measurement at $t_1 > t_2$ collapses the spin state of the {\em already decayed} $\mu^+$. This point of view was previously proposed for flavour entanglement in the $K^0 - \bar K^0$ system in Ref.~\cite{Bernabeu:2019gjs}, dubbing it as `post-tag'. A selection on the decay time of the longer-living kaon (`future') entails a selection on the decay time of the shorter-lived (`past'). Note, however, that the quantum nature of that decay-time correlation is yet to be established.\footnote{An illustrative example of non-quantum correlation is provided by top pair production in the dilepton decay channel at the Large Hadron Collider, $t \bar t \to \ell^+ \nu b \ell^- \nu b$. When considered over the full phase space, the $t \bar t$ pair is produced in a separable state. However, imposing a selection on the helicity angle $\cos \theta$ of the lepton $\ell$ from the longer-living top (`future') produces a selection on the analogous lepton angle for the shorter-living one (`past'). This effect is due to the spin correlation of the $t \bar t$ pair, and bears no relation to quantum entanglement. }

In our case, the hypothesis of Alice's measurement collapsing the $\mu^+$ spin state {\em in the past} agrees with all measurements Alice and Bob can perform on their respective muons --- the reason for it will become clear in the next section. 
Causality is not violated because a given set of momenta for the $\mu^+$ decay products can correspond to either $\mu^+$ spin: it is only after the post-selection based on Alice's measurement that the distributions corresponding to $|+\rangle$ and $|-\rangle$ spin states for $\mu^+$ are recovered. Moreover, causality {\em cannot} possibly be violated because the same outcome of measurements performed by Alice and Bob can be explained with a canonical past-to-future action, see section~\ref{sec:4.1}.

\section{Interpretation}

So far, we have considered both time orderings, $t_1 < t_2$ and $t_1 > t_2$, and computed the correlation between Alice’s and Bob’s measurements that would be expected in an actual experiment. We have used the postulates of quantum mechanics, most notably, the collapse of the state upon measurement --- in this respect, the $\mu^+$ decay also involves a measurement, namely of the momenta of its decay products~\cite{Aguilar-Saavedra:2024fig}. However, we have not made use of the spatial separation between Alice’s and Bob’s detectors. This distance is irrelevant in the present case, since the spin states remain stationary in the absence of external fields. Therefore, our results hold irrespectively of whether Alice’s and Bob’s measurements are space-like or time-like separated.

When the interval between Alice's and Bob's measurements is space-like, a paradoxical situation arises. As is well known, for space-like intervals the time ordering depends on the observer, according to special relativity. Thus, although there is no causal connection between the measurements, different observers may arrive at quite different interpretations of the same observations:
\begin{itemize}
\item An observer for whom $t_1 < t_2$ would argue that Alice's spin measurement of $\mu^-$ collapses the state; Bob's $\mu^+$ then has definite spin. Upon decay, the $e^+$ distribution reflects that spin eigenstate.
\item An observer for whom $t_1 > t_2$ would say that Bob's $\mu^+$ decay modifies the spin state of Alice's $\mu^-$, thereby influencing her spin measurement on a statistical basis. The post-selection based on Alice's results then recovers distributions {\em as if}\, Bob's $\mu^+$ had a definite spin.
\item A more unorthodox observer, still with $t_1 > t_2$, can sustain that Alice's $\mu^-$ spin measurement retroactively modified Bob's $\mu^+$ spin {\em before} it decayed.
\end{itemize}
A simpler experiment in which Alice and Bob both perform spin measurements in a space-like interval also admits (two) different interpretations for different observers. However, in the example discussed here the decay of $\mu^+$ turns the possible interpretations quite `asymmetric'.
And the three of them agree with the (expected) results of Alice and Bob's measurements --- as we have shown in sections \ref{sec:3} and \ref{sec:4}, the very same correlations are found either for $t_1 < t_2$ or $t_1 > t_2$. 

This disparity of interpretations of the same empirical observations reinforces the point of view that the Copenhagen interpretation of measurement, rather than describing the physical reality, is more a mathematical tool for calculations. The analysis of this (thought) experiment under alternative interpretations of QM might bring new light into the subject.

\section*{Acknowledgements}
This work has been supported by the Spanish Research Agency (Agencia Estatal de Investigaci\'on) through projects PID2022-142545NB-C21,  and CEX2020-001007-S funded by MCIN/AEI/ 10.13039/501100011033.

\end{document}